\documentclass[]{llncs}

\usepackage[english]{babel}
\usepackage[T1]{fontenc}
\usepackage[utf8]{inputenc}
\usepackage{lmodern}
\usepackage{url}
\usepackage{graphicx}
\graphicspath{{figs/}}
\usepackage{svg}
\usepackage{subcaption}
\usepackage{comment}

\usepackage{booktabs}
\usepackage{multirow}
\usepackage{tikz}
\usetikzlibrary{arrows,positioning, decorations.pathreplacing, shapes}

\usepackage{textcomp}
\usepackage{xcolor}
\usepackage{diagbox}

\usepackage{cite}

\newcommand{\ttt}[1]{\textbf{\texttt{\small{#1}}}}

\begin{document}
\title{Early Detection of Spam Domains with Passive DNS and SPF}
\author{}
\institute{}
%
%

\author{Simon Fernandez \and Maciej Korczy\'nski \and Andrzej Duda}

\authorrunning{}

\institute{Univ. Grenoble Alpes, CNRS, Grenoble INP, LIG, F-38000 Grenoble, France 
\email{\{first.last\}@univ-grenoble-alpes.fr}}

%
%
\maketitle              

\begin{abstract} 
Spam domains are sources of unsolicited 
mails and one of the primary vehicles for fraud and malicious activities such as phishing campaigns or malware distribution.
Spam domain detection is a race: as soon as the spam mails are sent, taking down the domain or blacklisting it is of relative use, as spammers have to register a new domain for their next campaign. 
To prevent 
malicious actors from sending mails, we need to detect them as fast as possible and, ideally, even before the campaign is launched.

In this paper, using near-real-time passive DNS data from Farsight Security, we monitor the DNS traffic of newly registered domains and the contents of their \ttt{TXT} records, in particular, the configuration of
the Sender Policy Framework, an anti-spoofing protocol for domain names and the first line of defense against devastating Business Email Compromise scams.
Because spammers and benign domains have different SPF rules and different traffic profiles, we build a new method to detect spam domains using features collected from passive DNS traffic. 

Using 
the SPF configuration and the traffic to the
\ttt{TXT} records of a domain, we accurately detect a significant proportion of spam domains with a low false positives rate demonstrating its potential in real-world deployments. 
Our classification 
scheme can detect spam domains before they send any mail, using only a single DNS query and later on, it can refine its classification by monitoring more traffic to the domain name. 

\end{abstract}

\section{Introduction}

For years, malicious mails have been representing a significant technical, economic, and social threat.
Besides increasing communication costs and clogging up mailboxes, malicious
mails may cause considerable harm by luring a user into following links to
phishing or malware distribution sites. 

Typically, malicious actors run campaigns with instant generation of a large number of mails.
Hence, their detection is a race: if we want to prevent their malicious activity, we need to detect 
spam domain names as soon as possible,~blacklist and block them (at the registration level). Once the campaign is over, domain blacklisting is
less effective because the recipients have already received  mails.

Early detection of spam domains that generate malicious mails is challenging.
One of the approaches is to leverage the Domain Name System (DNS) that maps domain
names to resource records that contain data like IP addresses. 
We can use DNS traffic and domain name characteristics to compute features for training and
running machine learning detection algorithms, even if malicious actors may try to hide
their traces and activities, and avoid domain takedown~\cite{holzMeasuringDetectingFastFlux, perdisciDetectingMaliciousFlux2009}. The main difference between 
various algorithms is the set of features used to train and run classifiers. The features mainly belong to four categories: i) lexical: domain names, randomness of characters, or similarity to brand names~\cite{antonakakisThrowAwayTrafficBots2012, bilgeExposurePassiveDNS2014, wangBreakingBadDetecting2015, korczynskiCybercrimeSunriseStatistical2018, lepochatSmorgasbordTyposExploring2019,antonakakisBuildingDynamicReputationb,lepochatPracticalApproachTaking2020}, ii) domain and IP address popularity: reputation systems based on diversity, origin of queries, or past malicious activity~\cite{kheirMentorPositiveDNS2014, haoPREDATORProactiveRecognition2016, lisonNeuralReputationModels2017, pochatTrancoResearchOrientedTop2019, antonakakisBuildingDynamicReputationb, lepochatPracticalApproachTaking2020, antonakakisDetectingMalwareDomainsa}), iii) DNS traffic: number of queries, their intensity, burst detection, or behavior changes~\cite{bilgeExposurePassiveDNS2014, lisonNeuralReputationModels2017}), and iv) WHOIS: domain registration patterns~\cite{lepochatPracticalApproachTaking2020, haoPREDATORProactiveRecognition2016, maroofiCOMARClassificationCompromised2020}.

In this paper, we propose a scheme for early detection of spam domains, even
before they send a single mail to a victim. 
It is based on the domain SPF (Sender Policy Framework) rules and traffic to the \ttt{TXT} records
containing them.

SPF rules 
are means for detecting forged sender addresses---they form the first line of defense in the case of, for instance, Business Email Compromise 
scams that represented over \$1.8 billion USD of losses in 2020~\cite{fbi_bec}. 
As malicious actors generally use newly registered domains for sending mails, they also configure the SPF rules for their domains to increase their reputation and thus avoid proactive detection. We have discovered that the content of the
SPF rules and traffic to the \ttt{TXT} records containing them are different for
malicious and benign domains. We have used these features to design a domain
classifier algorithm that can quickly detect spam domains 
based on passive
DNS traffic monitoring~\cite{farsightinc.FarsightSIE}. 
With low false positive rate 
and high true positive rate, our scheme can improve existing real-time systems for detecting and proactively blocking spam domains using passive DNS data.

The rest of the paper is organized as follows.  Section~\ref{sec:background} provides background on  
SPF and spam campaigns.
Section~\ref{sec:methodology} presents the proposed scheme.
Sections~\ref{sec:class} and \ref{sec:results} introduce the classification algorithms and present their results. 
We discuss other related approaches in
Section~\ref{sec:soa} and Section \ref{sec:conclusions}~concludes~the~paper.

\vspace{-0.2cm}
\section{Background}\label{sec:background}
\vspace{-0.2cm}

In this section, we 
describe the SPF protocol and the mail delivery process, highlighting the steps during which we gather features to detect malicious~activity.

\vspace{-0.3cm}
\subsection{Sender Policy Framework (SPF)}\label{sec:spf}

\begin{figure}[t]
\centering
\includegraphics[width=0.8\linewidth]{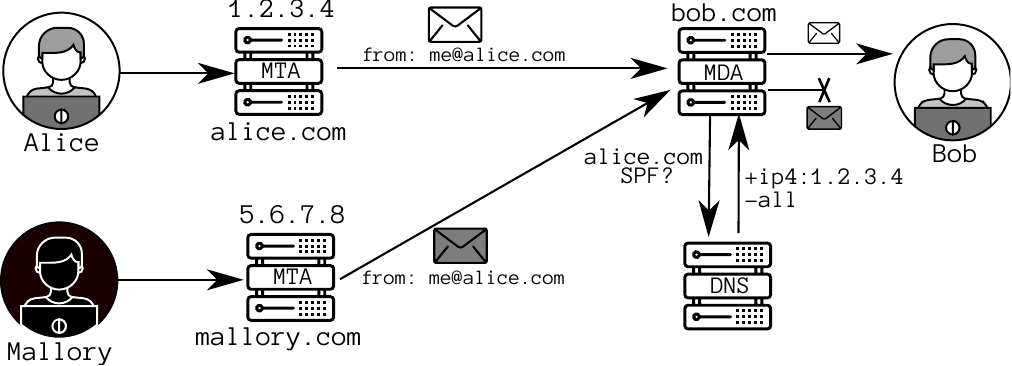}
\caption{Sending mails with SPF verification.}\label{fig:mail}
\vspace{-0.5cm}
\end{figure}

The Sender Policy Framework (SPF) \cite{kittermanSenderPolicyFramework2014a} is a protocol used to prevent  
domain (mail) spoofing.
Figure \ref{fig:mail} presents the procedure for sending mails and SPF verification.
Alice (sender) sends a benign mail to Bob (receiver). Mallory
(attacker) wants to send a mail that impersonates Alice to Bob. 
Mallory and Alice use their respective
servers (\ttt{mallory.com} and \ttt{alice.com}) to send mails. 

An effective anti-spoofing mechanism needs to differentiate the Mallory message
from the benign Alice mail. The current first lines of defense to protect users from spoofed mails include SPF
\cite{kittermanSenderPolicyFramework2014a},
DKIM~\cite{kucherawyDomainKeysIdentifiedMail2011}, and DMARC
\cite{kucherawyDomainbasedMessageAuthentication2015}. 
SPF is a set of text-form rules in \ttt{TXT} DNS resource records specifying a
list of servers allowed to send mails on behalf of a specific domain. During
mail delivery over the SMTP protocol, the recipient server authenticates the
sender Mail Transfer Agent (MTA) by comparing the given \ttt{MAIL FROM} (or
\ttt{HELO}) identity and the sender IP address with the content of the published SPF record.

In our example, the Mail Delivery Agent (MDA) on the Bob's server queries the DNS for a \ttt{TXT} record of the
sending domain (\ttt{alice.com}). This record contains the SPF rule of
\ttt{alice.com} and specifies which IP addresses can send mails on behalf of
this domain. The mail from Alice comes from a whitelisted server, so it gets
delivered. The Mallory's server was not whitelisted, so the (spoofed) mail is rejected.

A valid SPF version 1 record string must begin with \ttt{v=spf1} followed by
other SPF entries with the following structure: \ttt{<qualifier><mechanism>[:<target>]}.
%
The mail sender is matched with the \ttt{<mechanism>:<target>} part; the output is determined by the \ttt{<qualifier>}. 
Four types of \ttt{<qualifier>} are possible:
\ttt{PASS (+)} (the default mechanism), \ttt{NEUTRAL (\textasciitilde)},
\ttt{SOFTFAIL (?)}, \ttt{FAIL (-)}. 
The most common SFP mechanisms are the following:
\vspace{-0.2cm}
\begin{description}
    \item[\ttt{ip4, ip6}]-- the sender IP address matches the predefined IP address or the subnetwork prefix, 
    \item[\ttt{a, mx}]-- the domain has an \ttt{A} (or \ttt{MX}) record that resolves to the sender IP address,  
    \item[\ttt{ptr}]-- a verified reverse DNS query on the sender IP address matches the
      sending domain (not recommended by RFC 7208~\cite{kittermanSenderPolicyFramework2014a} since April 2014),
    \item[\ttt{exists}]-- the domain has an \ttt{A} record,
    \item[\ttt{include}]-- use the rules of another domain,
    \item[\ttt{all}]-- the default mechanism that always matches.
\end{description}
\vspace{-0.2cm}

To illustrate the operation of SPF rules, let us consider the following configuration for \ttt{example.com} domain:
\ttt{v=spf1 a ip4:192.0.2.0/24 -all} where the \ttt{A} record (\ttt{example.com A 198.51.100.1}) is stored in DNS. 
The SPF rule states that only 
a host with the IP address of
\ttt{198.51.100.1} (the \ttt{a} mechanism) or machines
in the \ttt{192.0.2.0/24} subnetwork (the \ttt{ip4}
mechanism) are permitted senders, all others are forbidden (the \ttt{-all} mechanism).

\vspace{-0.4cm}
\subsection{Life Cycle of a Spam Campaign}\label{sec:life_cycle}
\vspace{-0.1cm}

Most spam campaigns follow the same life cycle presented below. 
\vspace{-0.4cm}
\subsubsection{Domain registration.}
As most mail hosting companies deploy tools to 
prevent their users from sending spam, malicious actors need to register their own domains to send spam. 
To run multiple campaigns, spammers usually register
domains in bulk~\cite{haoUnderstandingDomainRegistration2013}.
Once the domains are registered, spammers  configure
zone files and fill the corresponding resource records in the DNS.
\vspace{-0.4cm}
\subsubsection{Configuration of anti-spoofing mechanisms.}
To use SPF, DMARC, or DKIM, each domain must have a \ttt{TXT} resource record
describing which hosts can send a mail on their behalf and deploying keys to
authenticate the sender. Even if DMARC is still not widely used, many benign domains
deploy SPF~\cite{maroofiDefensiveRegistrationSubdomain,9375477,CaseySPF}. Thus, a
mail from a domain without SPF configuration is likely to be flagged as spam (especially when combined with other indicators of malicious intent). 
To appear as benign as possible, spammers fill in at least the SPF rule in the \ttt{TXT} record. 
Our scheme extracts most of the features for detecting spam at this step because the
SPF records of spam domains are generally different from the configurations of
benign domains and even if a given domain has not yet sent a single mail, we can
access its SPF rules and detect suspicious configurations.
The SPF rules can be actively fetched by sending a \ttt{TXT} query to the domain (e.g., newly registered), but to avoid active scanning, we have chosen to use passive DNS to analyze 
\ttt{TXT} requests. 
In every detected spam campaign, we observe at least one \ttt{TXT} query that may originate from a spammer testing its infrastructure.

\begin{figure}
    \centering
    \vspace{-0.4cm}
    \includegraphics[width=.75\textwidth]{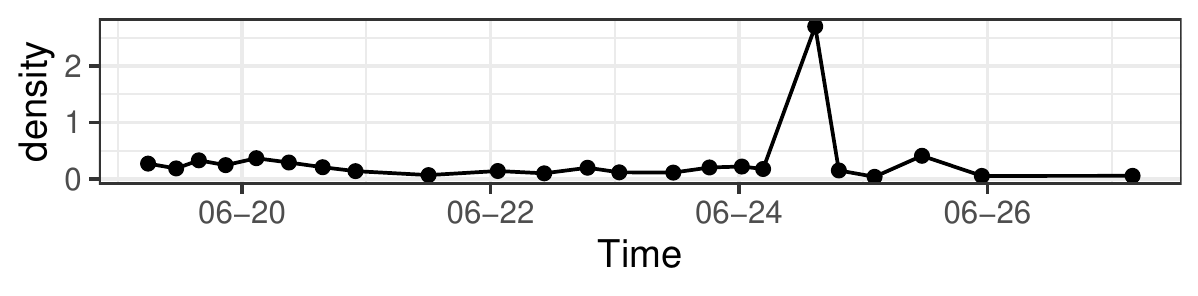}
\vspace{-0.3cm}
\caption{Density of DNS \ttt{TXT} traffic to a spam domain (\ttt{promotechmail.online})}\label{fig:spam_profile}
    \vspace{-0.6cm}
\end{figure}
\vspace{-0.5cm}
\subsubsection{Spam campaign.}
When a mail server receives a mail, it tries to resolve the \ttt{TXT} record
of the sending domain to get its SPF rule and checks for possible sender forgery. 
During a spam campaign, spammers send mails to many servers across the world. 
At the beginning of a campaign, the (validating) mail servers will all 
try to retrieve the \ttt{TXT} DNS record of the sender domain almost at the same time. Therefore, 
we expect to observe a surge in queries for \ttt{TXT} records.  
Figure~\ref{fig:spam_profile} presents traffic density (corresponding to the number of DNS queries over time, defined precisely later) to a spam domain detected during our study. 
The burst in the number of queries during a
time window of less than 24 h, then traffic dropping and never rising again is the typical profile of spammers.

\vspace{-0.4cm}
\subsubsection{Detection, blacklisting, and cleanup.}
When spam mails reach the targets, security experts and spam detection
algorithms parsing the mail content and its headers flag the sending domain as a spamming source and may report it to domain blacklists like SpamHaus \cite{spamhaus} or SURBL \cite{surbl}. 
When a domain appears on a blacklist, mail servers will likely drop mails from it. 
Future spam campaigns from this domain will be unsuccessful, so 
it becomes useless for spammers. Hosting services may also suspend the sending server whereas domain registrars may take down the spam domain as it often violates their terms of service and is considered as DNS abuse \cite{korczynskiCybercrimeSunriseStatistical2018,dnsabuse}. 
Once the domain is blacklisted (or taken down), spammers may just acquire another one 
and repeat the previous steps.

When looking for spammers, timing is the key: the sooner we detect a spamming domain,
the fewer mails it can send, and if an algorithm only detects a spam mail upon reception, it means that the 
campaign has 
started and reached some of the targets. 
This observation was the motivation for our scheme for early detection of spamming
domains even before the start of a spam campaign.
\vspace{-0.2cm}

\section{Scheme for Early Detection of Spam}\label{sec:methodology}
\vspace{-0.2cm}

In this section, we present the proposed scheme. 
It takes advantage of passive DNS data to obtain the SPF rules for a given domain and the frequency of the queries to retrieve them. 

\vspace{-0.4cm}

\subsection{Data Source: Passive DNS}\label{sec:pdns}


Passive DNS consists of monitoring DNS traffic by \emph{sensors} usually deployed above recursive resolvers to monitor
queries between a local resolver and authoritative name servers \cite{Weimer}. 
Locally observed queries are 
aggregated into feeds available for analyses. 
In this work, we have used the near-real-time Farsight SIE Passive DNS
channel 207~\cite{farsightinc.FarsightSIE} to obtain DNS traffic data for the \ttt{TXT} records and SPF rules for each domain.
We extract the following fields: the queried domain, the record type, the
answer from the authoritative server, a time window, and the number of times a given query was observed during the time window. 

To be effective, the scheme must analyze unencrypted DNS traffic. Therefore, it is not suitable when using the DNS over TLS (DoT) \cite{RFC-DoT} or DNS over HTTPS (DoH) \cite{RFC-DoH} standards that encrypt user DNS queries to prevent eavesdropping of domain names.  
To monitor such traffic, the scheme would have to be implemented, e.g., in public recursive resolvers providing DoT or DoH services.

\vspace{-0.4cm}

\subsection{Features Based on SPF Rules}\label{sec:features}

The SPF configuration for a given domain is stored in the \ttt{TXT} record of
the domain. 
Since most mail hosting services provide a default SPF records for their
customers, many domains share the same SPF rules. 
Nevertheless, some domains use custom SPF rules that whitelist specific servers. 
We have focused on the similarities of domains: two
domains that use the same custom SPF rules and whitelist the same IP addresses
are likely to be managed by the same entity. 
Therefore, if one domain starts
sending spam, it is reasonable to consider that the domains sharing
the same SPF rules are likely to be (future) spammers.

\begin{figure}
    \centering
    \vspace{-0.4cm}
    \includegraphics[width=0.7\textwidth]{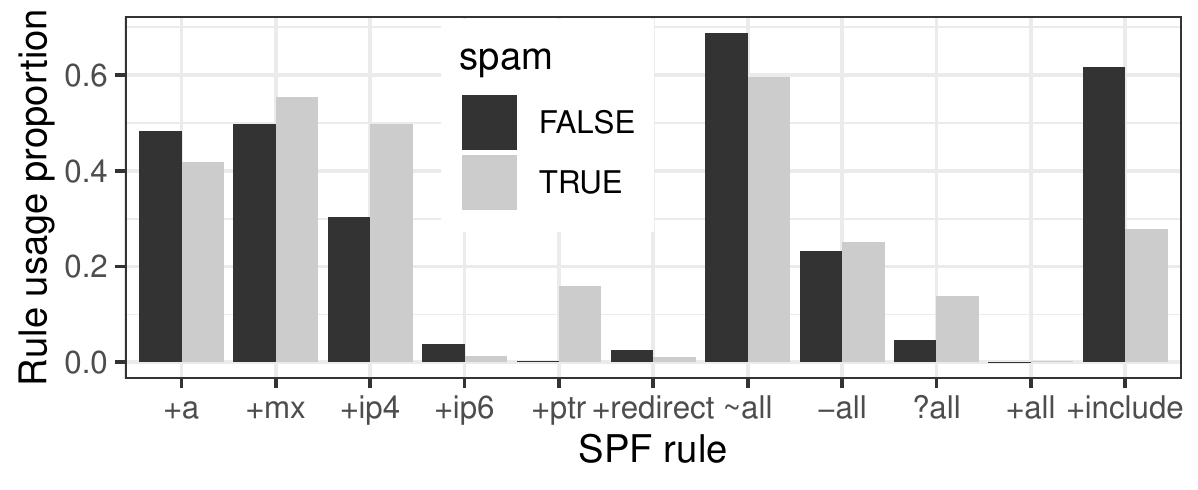}
    \vspace{-0.2cm}
    \caption{Usage proportion of SPF rules for benign and spamming domains}\label{fig:spf_count}
    \vspace{-0.5cm}
\end{figure}

We have analyzed the SPF configuration of spam and benign domains to see if they differ (we later discuss ground truth data in Section \ref{sec:ground_truth}). 
Figure~\ref{fig:spf_count} shows that benign and spam domains do not necessarily use the same rules. For example, benign domains more frequently use the \ttt{+include} mechanism while spammers \ttt{+ptr}. 
%

We presume that legitimate domains, hosted by major mail 
hosting providers, are more likely to have default configurations  
with the \ttt{+include} mechanism to indicate that a particular third party (e.g., a mail server of the provider) is authorized to send mails on behalf of all domains (e.g., in a shared hosting environment).
Spam domains 
may use custom mail servers instead, thus they are more likely to whitelist the IP addresses of their servers with, for instance, the \ttt{+ip4} mechanism. 
We suspect that in some cases spammers may not want to 
reveal the IP addresses of hosts sending spam. 
Therefore, they may use the \ttt{+all} mechanism (that accepts mails from all hosts) relatively more than legitimate domains whose administrators 
are concerned about rejecting spam mails from unauthorized host. Finally, the \ttt{+ptr} mechanism is marked as ``do not use'' since April 2014 by RFC 7208~\cite{kittermanSenderPolicyFramework2014a}. 
Major hosting providers seem to follow this recommendation, but individual spammers may not have changed their practices and continue to use this outdated but still supported mechanism.

For each domain, we compute the number of occurrences of each mechanism in its rule to generate the set of SPF features. Because not all possible combinations of qualifiers and mechanisms are actually used, we have selected the 
sets of qualifiers and mechanisms that appear in more than 0.1\% of domains to avoid overfitting, which leaves us 
the ones presented in Figure~\ref{fig:spf_count}.

\subsection{Graph Analysis of SPF Rules}\label{sec:graph}

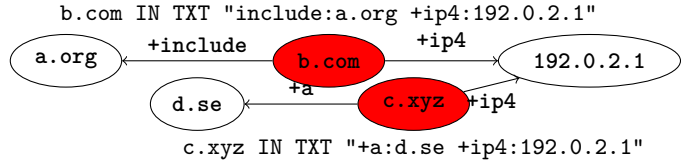
\begin{figure}
    \vspace{-0.3cm}
    \centering
    \begin{tikzpicture}
    \tikzstyle{node} = [draw, ellipse, minimum size=0.7cm]
    \tikzstyle{spam} = [fill=red]
    \tikzstyle{spf} = []

    \node[node] (a) {\ttt{a.org}};
    \node[node, spam] (b) [right = 2cm of a]{\ttt{b.com}};
    \node[node] (ip1) [right = 1.5cm of b]{\ttt{192.0.2.1}};
    \node[node,spam] (c) [below right = 0.1cm of b]{\ttt{c.xyz}};
    \node[node] (d) [left = 1.5cm of c]{\ttt{d.se}};

    \node[spf, anchor=south] (spf_b) at (b.north) {\texttt{b.com IN TXT ``include:a.org +ip4:192.0.2.1''}};
    \node[spf, anchor=north] (spf_c) at (c.south) {\texttt{c.xyz IN TXT ``+a:d.se +ip4:192.0.2.1''}};

    \draw[->] (b) -- node[anchor=south] {\ttt{+include}} (a);
    \draw[->] (b) -- node[anchor=south] {\ttt{+ip4}} (ip1);
    \draw[->] (c) -- node[anchor=north] {\ttt{+ip4}} (ip1);
    \draw[->] (c) -- node[anchor=south] {\ttt{+a}} (d);
\end{tikzpicture}
    \caption{Example of a relationship graph derived from SPF rules}\label{fig:graph_ex}
    \vspace{-0.4cm}
\end{figure}

Some SPF rules point to an IP address or a subnetwork prefix (like \ttt{ip4} and
\ttt{ip6}) and some point to domain names (like \ttt{include} and sometimes
\ttt{a} and \ttt{mx}). 
We build the relationship graph between domains and IP ranges as shown in Figure~\ref{fig:graph_ex}. 
For example, the edge between node A (\ttt{a.org}) and node B (\ttt{b.com}) reflects the fact that node B has an SPF rule that points to node A. 
The edge between \ttt{b.com} and \ttt{192.0.2.1} represents the fact that this IP address is used in the \ttt{+ip4} rule in the \ttt{b.com} SPF configuration.

This graph is built and updated in near real time: nodes and edges are added when domains with SPF data appear in the passive DNS feed, and spam domains (marked in red in Figure \ref{fig:graph_ex}) are 
added or deleted from blacklists (SpamHaus and SURBL in our scheme). 
Thus, over time, the graph becomes more complete, providing more precise relationships and features for domain classification.

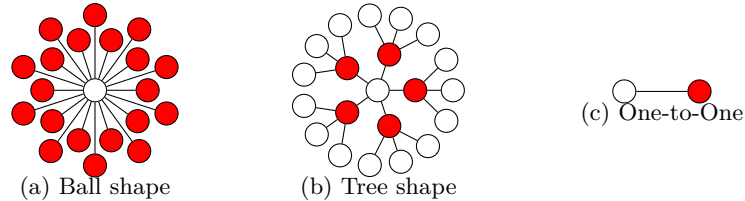
\begin{figure}[h]
    \begin{center}
    \begin{subfigure}{0.3\textwidth}
        \begin{center}
        \begin{tikzpicture}
    \tikzstyle{node} = [draw, circle, minimum size=0.2cm]

    \node[node] (center) {};

    \def \count {10}
    \def \radius {0.7}

    \foreach \s in {1,...,\count}
    {
    \node[node, fill=red] (inner-\s) at ({360/\count * (\s - 1)}:\radius) {};
    \draw (inner-\s) -- (center);
    }
    
    \def \radius {1}
    \foreach \s in {1,...,\count}
    {
    \node[node, fill=red] (outer-\s) at ({360/\count * (\s - 1) + (360/(2*\count))}:\radius) {};
    \draw (outer-\s) -- (center);
    }

\end{tikzpicture}
        \end{center}
        \vspace{-1cm}
        \caption{Ball shape}\label{fig:spf_ball}
    \end{subfigure}
    \begin{subfigure}{0.3\textwidth}
        \begin{center}
        \begin{tikzpicture}
    \tikzstyle{node} = [draw, circle, minimum size=0.2cm]

    \node[node] (center) {};

    \def \count {5}
    \def \radius {0.5}

    \foreach \s in {1,...,\count}
    {
    \node[node, fill=red] (inner-\s) at ({360/\count * (\s - 1)}:\radius) {};
    \draw (inner-\s) -- (center);
    }
    
    \def \radius {1}
    \def \leafs {3}
    
    \foreach \s in {1,...,\count}
    {
        \foreach \n in {1,...,\leafs}{
            \node[node] (outer-\s\n) at ({(360/\count * (\s - 1) + 360/(3*\count) * (\n - 0.5 -\leafs/2))}:\radius) {};
            \draw (outer-\s\n) -- (inner-\s);
        }
    }

\end{tikzpicture}
        \end{center}
        \vspace{-1cm}
        \caption{Tree shape}\label{fig:spf_tree}
    \end{subfigure}
    \begin{subfigure}{0.3\textwidth}
        \begin{center}
        \begin{tikzpicture}
    \tikzstyle{node} = [draw, circle, minimum size=0.2cm]

    \node[node] (center) {};
    \node[node, fill=red] (spam) [right of=center] {};
    \draw (center) -- (spam);
\end{tikzpicture}
        \end{center}
        \vspace{-1cm}
        \caption{One-to-One}\label{fig:spf_one}
    \end{subfigure}
    \end{center}
    \vspace{-0.3cm}
    \caption{SPF relation graph for spam domains}\label{fig:spf_graph}
    \vspace{-0.5cm}
\end{figure}

We have analyzed different structures in the graph built from our dataset and
detected distinctive patterns. 
Figure~\ref{fig:spf_graph} shows three examples of the observed structure types to illustrate some typical SPF configuration relationship graphs for spam domains.
Red nodes represent spamming domains and white nodes correspond to the targets of their
SPF rules. 
Figure~\ref{fig:spf_ball} shows the pattern in which multiple spam domains share the
same configuration: they have a rule targeting the same IPv6 network (these domains are likely to be managed by the same entity).
Figure~\ref{fig:spf_tree} presents spam domains that have an \ttt{include}
mechanism that points to the same domain and exactly three other custom targets
that no other domain uses (this is the case when domains are hosted by a hosting provider that provides an SPF configuration for inclusion by its clients). Finally, many spam domains have rules like in
Figure~\ref{fig:spf_one} in which a domain has a single target (a custom IP
address) that no other domain uses.
\vspace{-1cm}

The study of these structures can highlight potential spam domains. In our
dataset, we found structures like in Figure~\ref{fig:spf_ball} or
Figure~\ref{fig:spf_tree} in which dozens of domains used the same rule and the
majority of them appeared on spam blacklists. 
As such, it
is reasonable to assume that the remaining domains are likely to have not yet been detected 
or are not yet active spam domains. 

To detect the structures indicating spam domains, we have defined two unique
features describing the properties of domains in the relationship graph.

\vspace{-0.4cm}
\subsubsection{Toxicity.}
We define the \emph{toxicity} of a node as the proportion of its neighbors that 
are flagged as spam in the graph, or $1$ if the domain itself is flagged~as~spam.
With this
metric, SPF targets used by known spammers get a high value of \emph{toxicity}. 
To detect the domains that use rules with high \emph{toxicity} targets, we
compute the \emph{Max Neighbor Toxicity}: the maximum \emph{toxicity}
amongst all the targets of a domain. This way, if a domain has a target mainly
used by spammers, its \emph{Max Neighbor Toxicity} is high.

\vspace{-0.4cm}
\subsubsection{Neighbor Degree.}
For each node, we look at the degrees of its neighbors: is it connected to
highly used domains and IP addresses? Or, is it using custom targets that no other domain
uses? We expect spamming domains to more likely use custom targets that no
other domains use (with a small degree in the graph) like in
Figure~\ref{fig:spf_one}, compared to benign domains that would use the
default configurations of the hosting service  and share the same targets as many other
domains (with a high degree in the graph). 

\begin{figure}
    \begin{center}
      \vspace{-0.4cm}
    \includegraphics[width=.75\textwidth]{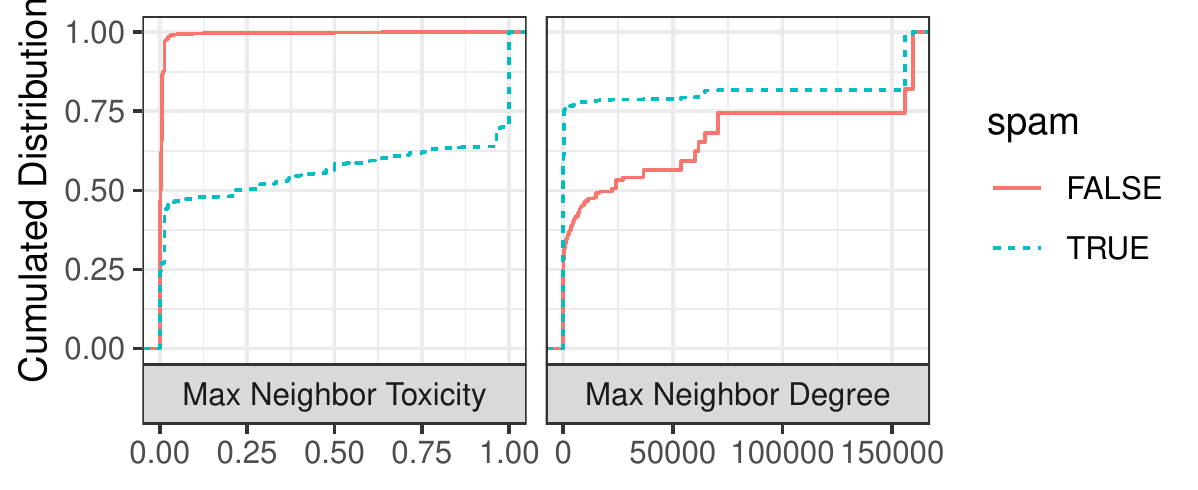}
   \vspace{-0.2cm}
    \caption{Cumulative distributions of Max Neighbor Toxicity and Max Neighbor Degree for spamming and benign domains.}\label{fig:max_deg_tox}
    \end{center}
      \vspace{-0.8cm}
\end{figure}

Figure~\ref{fig:max_deg_tox} shows that the expected differences of \emph{Max
Neighbor Toxicity} and \emph{Max Neighbor Degree} between spammers and benign 
domains match our hypothesis: spammers are more likely to use targets shared by
some 
other spammers and are more likely to use custom targets with low degrees in the graph.

\vspace{-0.2cm}
\subsection{Time Analysis of Traffic to DNS \ttt{TXT} Records}
\vspace{-0.3cm}

When a domain starts a spam campaign, we expect 
multiple servers to query DNS
for the \ttt{TXT} record of the sender domain to check its SPF configuration. Therefore, we can observe an unusual number of queries related to the (newly registered)~domain.
The passive DNS feed we use contains aggregated queries over a given time
window: when a DNS query is detected by a sensor, it is inserted in an
aggregation buffer with the insertion timestamp. The subsequent identical
queries only increase a counter in the buffer. When the buffer is full, the
oldest inserted queries are flushed out, yielding an aggregated message with
the
query, the answer from the authoritative server, and three extra fields:
\ttt{time\_first, time\_last}, and \ttt{count} meaning that the query was
seen \ttt{count} times during the time window from \ttt{time\_first} to
\ttt{time\_last}. 

From these aggregated messages, we compute the traffic density by dividing the number of queries (in the \ttt{count} field) by the window duration, and then, dividing this value by the time between the end of the window and the end of the previous window to take into account the time windows in which there is no traffic. The resulting formula is the following:
\[density(i) = \frac{\ttt{count}}{\ttt{time\_last} - \ttt{time\_first}} \times \frac{1}{message\_end(i) - message\_end(i-1)}\]
For a more in-depth definition of the density and an explanation on how we handled overlapping windows, see Appendix~\ref{app:density}.

\vspace{-0.4cm}
\subsubsection{Max Variation.}
To detect large variations in density, we compute the \emph{Max
Variation} feature defined as the maximum density variation during 
24 h. Domains with a slowly increasing traffic have a low \emph{Max
Variation} and those with a spike in the number of \ttt{TXT} queries, a high \emph{Max Variation}. We compute two versions of this feature: i) the \emph{Global Max
Variation}, using the same time steps to compare all domains and ii) the \emph{Local Max Variation} in which a custom time step is computed for each domain. See Appendix~\ref{app:density} for more details about the difference between these features.

\begin{figure}
    \begin{center}
       \vspace{-0.4cm}
    \includegraphics[width=0.75\textwidth]{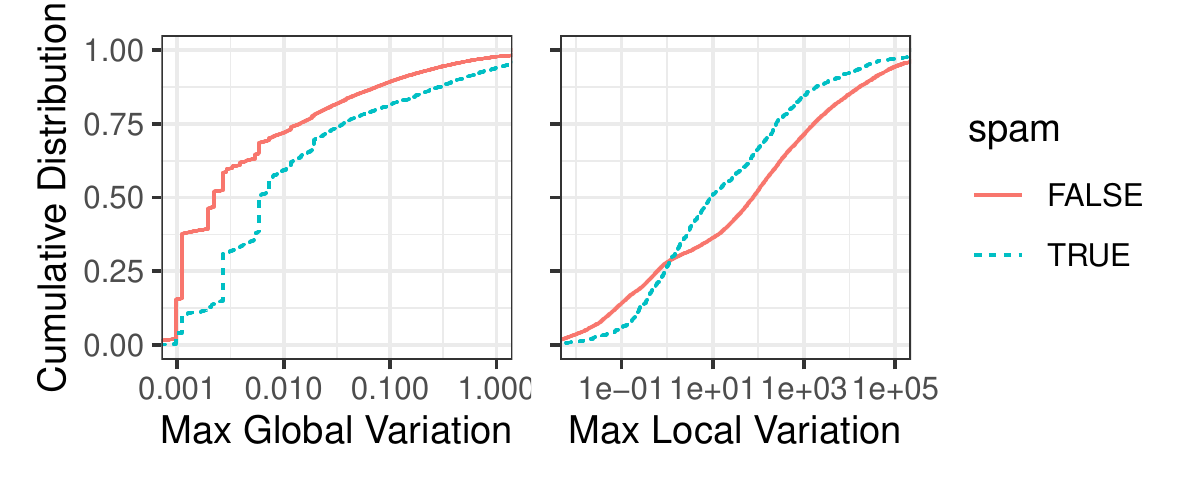}
       \vspace{-0.2cm}
    \caption{Cumulative distribution of Max Variation (log scale x-axis)}\label{fig:max_var}
    \end{center}
       \vspace{-0.4cm}
\end{figure}

\vspace{-0.5cm}
Figure~\ref{fig:max_var} presents the cumulative distribution of the two features. As expected, we observe that spam domains have a relatively higher \emph{Max Global Variation} when all domains share the same time steps.


However, when we look at the \emph{Max Local Variation}, we observe that benign domains tend to have a higher variation.
The distributions are different because this feature is close to the average density variation: domains with a lot of traffic variation and small windows will have a higher \emph{Local Variation}, whereas spam domains with almost no traffic except for a few spikes will have a lower \emph{Local Variation} due to long periods of inactivity before a spike.

\vspace{-0.4cm}
\section{Classifiers}\label{sec:class}
\vspace{-0.2cm}

In this section, we present the classifiers used for the detection of spam based
on the proposed features.
\vspace{-0.4cm}

\subsection{Ground Truth}\label{sec:ground_truth}

We have taken the precaution of carefully selecting the domains
in our ground truth.
We recorded four
months (between May and August 2021) of passive DNS traffic to \ttt{TXT} records from Farsight Security~\cite{farsightinc.FarsightSIE}. 
Because most spam domains are newly registered and discarded
as soon as they are blacklisted, we only considered newly registered domains.
From the ICANN Central Zone Data Service (CZDS) \cite{icann-czds}, we have built a list of new domains by
computing the difference between consecutive versions of each generic Top Level
Domain (gTLD) zone files.
Appendix \ref{sec:datastats} provides the general statistics of the collected dataset.

Using SURBL \cite{surbl} and SpamHaus \cite{spamhaus} spam blacklists, we have identified all domains (in near-real time) in our database flagged by one of these sources. 
Spam blacklists are not perfect and sometimes they may flag benign domains
as spam.  Therefore, to obtain reliable ground truth, we added an extra layer of verification: a domain is labeled as
\vspace{-0.1cm}
\begin{itemize}
\item \textbf{benign} if it has not been blacklisted and has been active during the entire period of the study (and has a valid \ttt{A} and \ttt{NS} records), or
\item \textbf{malicious} if it was blacklisted by SURLB or SpamHaus and was 
taken~down.

\end{itemize}
\vspace{-0.1cm}



With these criteria, our ground truth dataset contained 37,832 non-spam and 2,392 spam domains. 

\vspace{-0.2cm}
\subsection{Classifier}\label{sec:classifier}

For spam detection, 
it is crucial to keep the True
Negative\footnote{True
Negative: non-spam domain correctly classified as such} Rate (TPR) as high as possible to avoid flagging benign domains as spam.
Once a True Negative Rate of at least 99\% is achieved, we maximize the True Positive\footnote{True Positive: spam domain correctly classified as malicious} Rate (TPR) to detect as many
spam domains as possible.
%
%
To compare classification results we use true negative and true positive rates, and the F1-score as described in Appendix \ref{app:classifier_metrics}.
We explored multiple classifiers and parameters with Weka \cite{weka}, 
then implemented two of them with the \ttt{scikit-learn} \cite{scikit-learn}
Python library, for better benchmarking. Two classifiers that performed the best are:
\vspace{-0.2cm}
\begin{description}
\item[C4.5 or J48:] a decision tree able to describe non-linear relations between
  features. It highlights complex conditional relations between features.
\item[Random Forest:] a set of multiple decision trees with a voting system to
  combine their results. Its drawback is low explainability.
\end{description} 

\begin{table}[h!]
    \vspace{-1cm}
    \caption{Features used by the classifiers}\label{tab:properties}
    \begin{center}
    \begin{tabular}{clccc}
    \toprule
    Category & Feature & Outcome \\
    \midrule
    \multirow{4}*{SPF Rules}& Number of\ldots &\\
     & 
     \ttt{+all, +mx, +ptr, -all} & Malicious \\
     & 
     \ttt{+a, +include, +redirect, \textasciitilde all} & Benign \\
     & 
     \ttt{+ip4, +ip6, ?all} & Mixed\footnotemark \\
    \midrule
    \multirow{2}*{SPF Graph}& Max Neighbor Degree & Benign \\
    & Max Neighbor Toxicity & Malicious \\
    \midrule
    \multirow{2}*{Time Analysis}& Max Global Variation & Malicious \\
    & Max Local Variation & Benign \\
    \toprule
    \end{tabular}
    \end{center}
    \vspace{-0.7cm}
\end{table}

\footnotetext{Depends on how many times the rule is present in the configuration}

%
We use the k-fold cross-validation technique with $k$ set to 5 (see Appendix~\ref{app:classifier_metrics} for more information).
The number of spam domains in our ground truth dataset represents less than
10\% of all domains. 
The decision tree algorithms are not suitable for classification problems with a skewed class distribution.
Therefore, we have used a standard class weight algorithm for processing imbalanced data \cite{zhu2018class}
implemented in the \ttt{scikit-learn} Python library~\cite{scikit-learn}.

Table~\ref{tab:properties} summarizes the features used by the classifiers and whether they indicate maliciousness or benignness of the domain.

\section{Classification Results}\label{sec:results}

We evaluate the efficiency of the classifiers with two sets of features: i) the \textit{static} set without the time analysis features (Max Variation) and ii) the \textit{static + dynamic} set that includes both static and the time analysis features. 
We have distinguished between the sets because
even if the efficiency is lower without the time analysis features, we can get the static
features (SPF configuration and graph properties) from a single
\ttt{TXT} query to the target domain allowing for a rapid detection of most spam domains. 
%
Then, we can refine the classification by adding the time based features that are more robust against evasion techniques but require more time to detect spam domains.

\vspace{-0.2cm}
\subsection{Performance Evaluation}
\vspace{-0.2cm}

\begin{figure}
\begin{center}
\vspace{-0.4cm}
    \includegraphics[width=.7\textwidth]{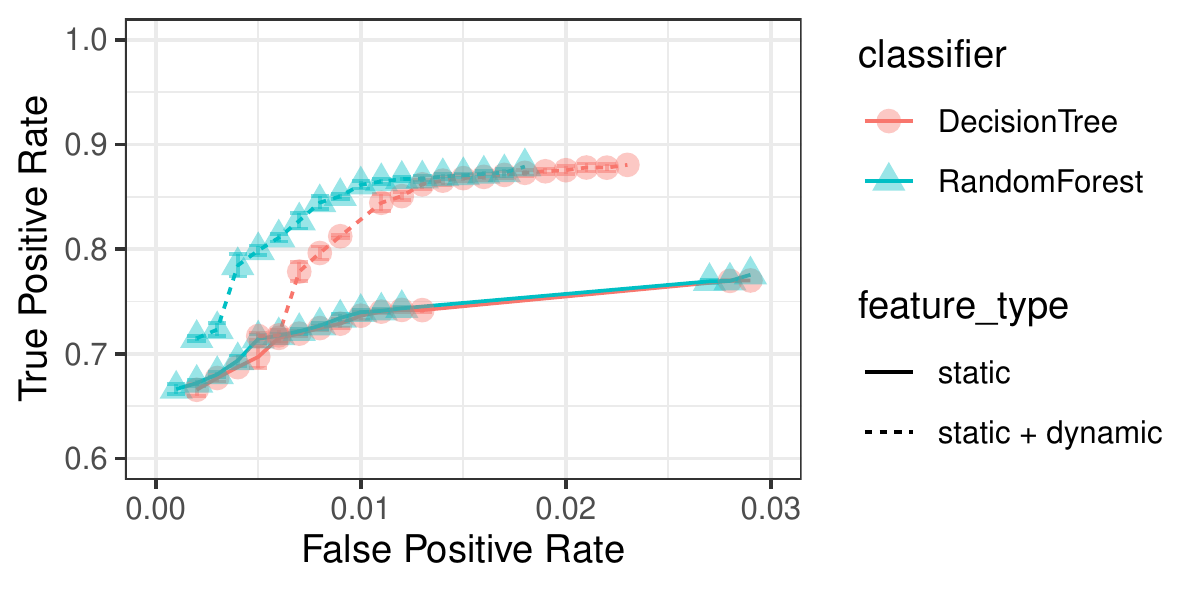}
    \caption{ROC curve for different classifiers on two sets of features}\label{fig:roc}
\end{center}
\vspace{-1cm}
\end{figure}

Figure~\ref{fig:roc} compares the Receiver Operating Characteristic (ROC) curves of each classifier for two sets of
features (to see better the  
differences in performance, we zoom into high values of TPR). 
When training the classifiers, we change the weight of the spam class to change the reward of accurately finding a spam domain. If the spam class weight is low, the classifier will be less likely to risk getting a false positive. On the contrary, if the spam class weight is high, the classifier gets higher reward if it accurately flags a spam domain. Therefore, the classifier will ``take more risks'', reducing its 
TNR to increase TPR.
If we require the False Positive Rate (benign domains flagged as spam) under 1\%,
the Random Forest is the best algorithm reaching a True Positive 
Rate of 74\% using only the static set and 85\% once we add the time analysis features.

\begin{figure}
\begin{center}
\vspace{-0.4cm}
    \includegraphics[width=.7\textwidth]{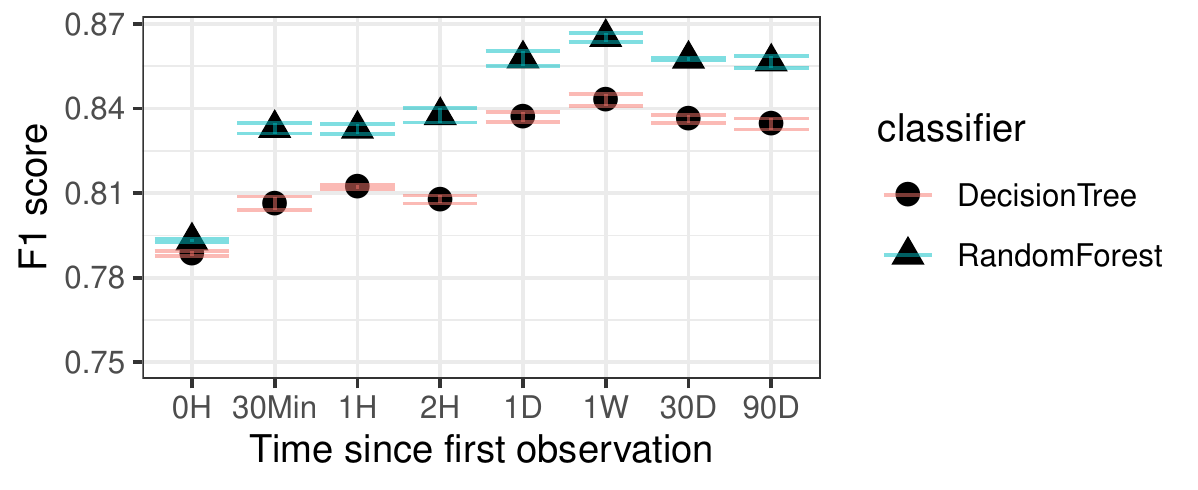}
    \caption{F1-score of classifiers after the first appearance of each domain}\label{fig:time_detection}
\end{center}
\vspace{-0.8cm}
\end{figure}

Figure~\ref{fig:time_detection} illustrates how long we need to monitor a domain
so that the classifiers reach their best efficiency. Over time, we observe traffic to
each domain and the time analysis features get more precise (until one week), which improves
classification. 
Both classifiers reach almost the best detection performance (computed as the F1-score) after observing a domain for one day.
%

\vspace{-0.4cm}
\subsection{Detection Time}

The static results (labeled as 0H in Figure~\ref{fig:time_detection}) show the efficiency of the scheme when a single \ttt{TXT} request is observed. In this case, the classifier has no time properties of the traffic and only uses the static 
features (SPF Rules and SPF Graph). 
We can replace passive detection of SPF Rules with active DNS scans 
%
(assuming we have a list of newly registered domain names, which is generally the case for legacy and new gTLDs but not for the vast majority of ccTLD \cite{Reputation,dnsabuse}): by actively querying the \ttt{TXT} records of new domains and classifying them based on their SPF configuration and formed relationships. Then, 
over time, as we passively observe traffic to the domain records, the performance of the classifier improves achieving very good results after 30 minutes (F1-score of 0.83) of monitoring (with Random Forest) in comparison with the F1-score of 0.86 after one day.

\begin{figure}
\begin{center}
\vspace{-0.6cm}
    \includegraphics[width=0.9\textwidth]{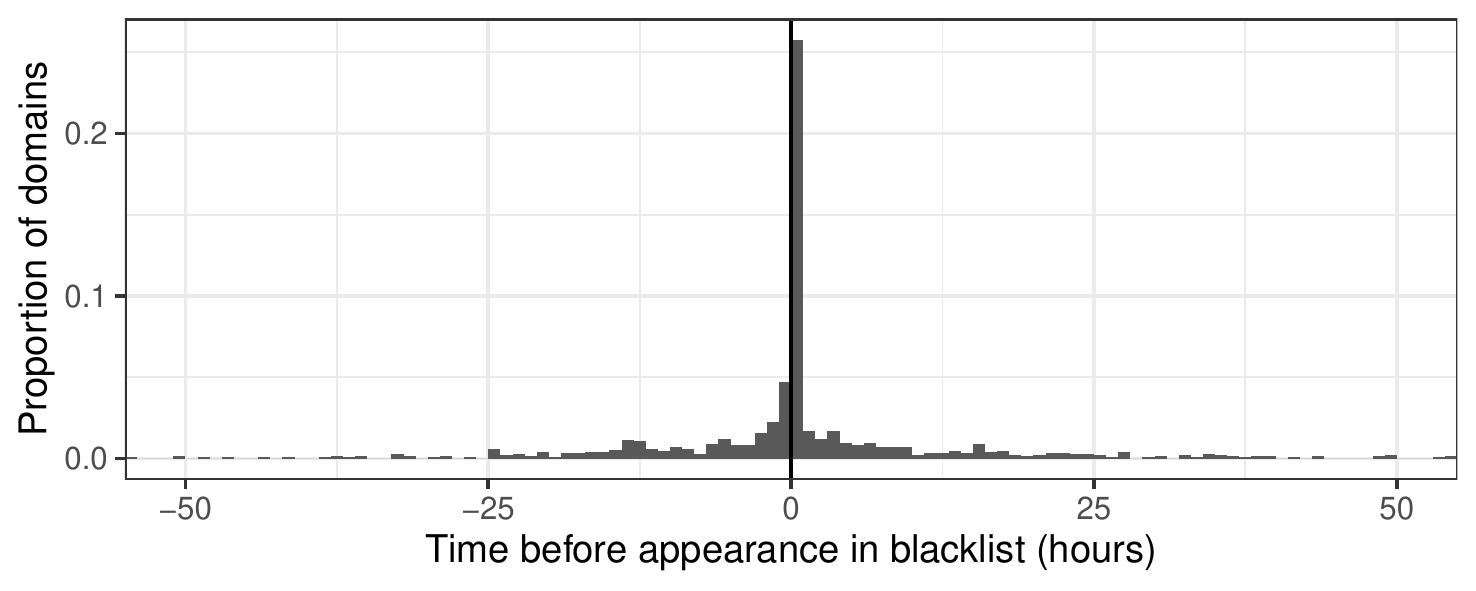}
\vspace{-0.2cm}
    \caption{Time before detected spam domains appear in commercial blacklists}\label{fig:detection_delay}
\end{center}
\vspace{-0.8cm}
\end{figure}

Using only static features, 
we compared the spam domain detection speed of our scheme with two commercial blacklists (SpamHaus and SURBL). In Figure~\ref{fig:detection_delay}, we plotted the time elapsed between the detection by our scheme and the appearance of  domains in the blacklists (with an hourly granularity). We limited the graph to 50 hours, 
but considerable number of domains only appear in the commercial blacklists weeks after we detect them. Positive values mean that our scheme was faster: for 70\% of the detected spam domains, our scheme was faster than the commercial blacklists. However, 26\% of the domains detected by our scheme appear in the commercial blacklists in the following hour, whereas 30\% of the domains are detected more than 24 hours in advance. The negative values represent domain names where our scheme was slower than the commercial blacklists: 30\% of the domains were already in the blacklists when they were observed in our passive DNS feed for the first time and classified as spam.

\vspace{-0.2cm}
\subsection{Feature Importance}
\vspace{-0.2cm}

The importance of each feature was computed by looking at how selective the feature was in the Random Forest classifier~\cite{scikit-learn}. The importance of each feature and each category is described in Table~\ref{fig:feature_importance}. 
It is not a surprise that the Maximum Neighbor Toxicity is by far the most important feature: a domain whitelisting the same IP addresses and domains as a known spamming domain is very likely to be managed by spammers. The most important SPF rule for classification is \ttt{+ptr}: as we discussed in Section~\ref{sec:features}, this rule is almost never used by benign domains (following the RFC 7208 recommendations). Lastly, the Global Max Variation is the most important dynamic feature: massive increases in the number of queries to a domain is a distinctive trait of spamming domains, as presented in Section~\ref{sec:life_cycle}, but this feature is only useful after the start~of~the~spam~campaign.

\vspace{-0.5cm}
\begin{table}
    \centering
    \caption{Importance of each feature for the Random Forest classifier}
    \begin{tabular}{ll}
        Feature & Importance \\
        \midrule        
        \textbf{
        SPF Graph features} & \textbf{0.574515} \\
        \hspace{4mm}\ttt{neighbor\_max\_toxicity} & 0.463689 \\
        \hspace{4mm}\ttt{neighbor\_max\_degree} & 0.110826 \\
        \midrule
        \textbf{
        SPF Rules features} & \textbf{0.232846} \\
        \hspace{4mm}\ttt{+ptr} & 0.100481 \\
        \hspace{4mm}\ttt{+a} & 0.029005 \\
        \hspace{4mm}\ttt{+ip4} & 0.028789 \\
        \hspace{4mm}\ttt{+mx} & 0.021006 \\
        \hspace{4mm}\ttt{+include} & 0.017561 \\
        \hspace{4mm}\ttt{?all} & 0.013728 \\
        \hspace{4mm}\ttt{\textasciitilde all} & 0.011522 \\
        \hspace{4mm}Other rules & < 0.01 \\
        \midrule
        \textbf{
        Time Analysis features} & \textbf{0.192638} \\
        \hspace{4mm}\ttt{global\_max\_variation\_24h} & 0.122167 \\
        \hspace{4mm}\ttt{local\_max\_variation\_24h} & 0.036828 \\
        \hspace{4mm}\ttt{global\_max\_triggers\_24h} & 0.022380 \\
        \hspace{4mm}\ttt{local\_max\_triggers\_24h} & 0.011263 \\
    \end{tabular}
    \label{fig:feature_importance}
    \vspace{-0.7cm}
\end{table}

\vspace{-0.3cm}
\section{Related Work}\label{sec:soa}

The four main  categories of features used to detect malicious domains are the following: i) Lexical: domain name, randomness of characters, or similarity to brand names~\cite{bilgeExposurePassiveDNS2014, lepochatSmorgasbordTyposExploring2019,antonakakisBuildingDynamicReputationb,lepochatPracticalApproachTaking2020, maroofiCOMARClassificationCompromised2020}, ii) Domain and IP address popularity: reputation systems based on diversity, origin of queries, or past malicious activity~\cite{haoPREDATORProactiveRecognition2016, pochatTrancoResearchOrientedTop2019, antonakakisBuildingDynamicReputationb, lepochatPracticalApproachTaking2020, antonakakisDetectingMalwareDomainsa, maroofiCOMARClassificationCompromised2020}, iii) DNS traffic: number of queries, intensity, burst detection, behavior changes~\cite{bilgeExposurePassiveDNS2014, lisonNeuralReputationModels2017},
iv) WHOIS (domain registration data): who registered a given domain\footnote{not available after the introduction of the General Data Protection Regulation (GDPR) and the ICANN Temporary Specification \cite{ICANNTemp}.}, when, and at which registrar~\cite{lepochatPracticalApproachTaking2020, haoPREDATORProactiveRecognition2016, maroofiCOMARClassificationCompromised2020}.
Other methods develop specific features extracted from the content of mails: size of the mail, links, or redirections~\cite{marchalOfftheHookEfficientUsable2017,maroofiCOMARClassificationCompromised2020}.
With the selected features, machine learning algorithms classify malicious and benign domains.

With respect to the methods that work on passive data such as Exposure~\cite{bilgeExposurePassiveDNS2014} that need some time to detect abnormal or malicious patterns, we focus on early detection of spam domains. 
Exposure for instance, needs around a week of observation before possible detection, while we achieve a F1-score of 79\% based on a single DNS query.    
Our scheme can be applied at early stages of a domain life cycle: using passive (or active) DNS, we can obtain SPF rules for newly registered domains and classify them immediately, or wait until we detect \ttt{TXT} queries to that domain and refine the classification using hard-to-evade temporal~features.

Other methods generally try to detect abnormal or malicious patterns at
later phases of the domain life cycle. Schemes based on content or long period traffic analysis may reach high efficiency but generally cannot run before or at the beginning of an attack. 
Schemes using lexical and popularity features can run preemptively but may have reduced efficiency, compared to dynamic schemes.

Our scheme may complement other approaches that aim at detecting spam during other phases in the life cycle of spam campaigns and other algorithms that rely on a variety of different features. 

\vspace{-0.2cm}
\section{Conclusion}\label{sec:conclusions}
\vspace{-0.2cm}

In this paper, we have proposed a new scheme for early detection of spam domains based on the content of domain SPF 
rules and traffic to the \ttt{TXT} records containing them.
With this set of features, our best classifier detects 85\% of 
spam domains while keeping a False Positive Rate under 1\%. 
The detection results are remarkable given that the classification only uses the
content of the domain SPF rules 
and their relationships, and hard to evade features based on DNS traffic. 
The performance of the classifiers stays high, even if they are only given the
static features that can be gathered from a single \ttt{TXT} query (observed passively or actively queried).

With a single request to the \ttt{TXT} record, we detect 75\% of the spam domains, possibly before the start of the spam campaign. 
Thus, our scheme brings important speed of reaction: we can detect spammers with good performance even before any mail is sent and before a spike in the DNS traffic. 
To evaluate the efficiency of the proposed approach based on passive DNS, we did not combine the proposed features with
other ones used in previous work like 
domain registration patterns~\cite{lepochatPracticalApproachTaking2020, haoPREDATORProactiveRecognition2016, maroofiCOMARClassificationCompromised2020}.
In practical deployments, the classification can be improved by adding other 
features based on, e.g., the content of potentially malicious mails or the lexical patterns of the domain names.



The features used in our scheme yield promising results, so adding them to
existing spam detection systems will increase their 
performance 
without large computation overhead as SPF data can easily be extracted
from near-real-time passive DNS feeds already used in some schemes.

\section*{Acknowledgements}
We thank Paul Vixie and Joe St Sauver (Farsight Security), the reviewers, our shepherd, and Sourena Maroofi for their valuable and constructive feedback.
We thank Farsight Security for providing access to the passive DNS traffic as well as SpamHaus and SURBL for the spam blacklists. This work was  partially supported by the Grenoble Alpes Cybersecurity Institute under
contract ANR-15-IDEX-02 and by the DiNS project under contract ANR-19-CE25-0009-01.

\bibliographystyle{splncs04}
\bibliography{spf2,article,soa}

\section*{Appendix}
\appendix

\section{Density Computation}\label{app:density}
Comparing the time windows of multiple domains in passive DNS data is a complex task: each
window has a different size and we have no information on how the queries are
spread inside it. 

\begin{figure}    
\vspace{-0.6cm}
    \begin{center}
    \begin{tikzpicture}
    \draw[thick, ->] (0,0) -- (7,0) node[font=\scriptsize,below left=3pt and -8pt]{time};
    \foreach \x in {0,...,6}
        \draw (\x,2pt) -- (\x cm,-2pt) node[anchor=north] {$\x$};

    \draw[red, |-|, dotted] (0,0.5) -- node[anchor=north]{10} (2,0.5);
    \draw[red, |-|, dotted] (1,0.7) -- node[anchor=south]{2} (3,0.7);
    \draw[blue, |-|] (2,0.5) -- node[anchor=north]{5} (3,0.5);

    \draw[red, |-|, dotted] (3,0.5) -- node[anchor=north]{3} (6,0.5);
    \draw[blue, |-|] (3,0.7) -- node[anchor=south]{5} (4,0.7);
    \draw[blue, |-|] (5,0.7) -- node[anchor=south]{7} (6,0.7);

    \draw[dotted, thick] (0, 0) -- (0, 1.5) node[anchor=south]{flush 0};
    \draw[dotted, thick] (3, 0) -- (3, 1.5) node[anchor=south]{flush 1};
    \draw[dotted, thick] (6, 0) -- (6, 1.5) node[anchor=south]{flush 2};    
    
    \draw[dotted, thick, blue] (4, 0) -- (4, 1) node[anchor=south]{B flush};  

    \draw[red, |-|, dotted] (7.5,0.8) -- node[anchor=west]{\hspace{0.3cm}Domain A} (8,0.8);
    \draw[blue, |-|] (7.5,0.4) -- node[anchor=west]{\hspace{0.3cm}Domain B} (8,0.4);

\end{tikzpicture}
    \vspace{-0.6cm}
    \caption{Computation of traffic density from Passive DNS messages}\label{fig:merge_windows}
    \end{center}
    \vspace{-0.6cm}
\end{figure}



The query density of multiple domains can only be compared if they are computed the same way, over the same time period. 
If a period starts or ends in the middle of a domain time window, we need to make an assumption about how the queries are
spread inside the time window, to determine how many queries are inside the time period. However, we do not have such 
information so a period can only start and end at a timestamp that it is not included in any time 
window. We call those usable timestamps \emph{flushes}. Then, the query density 
of a domain between two flushes is computed by measuring the time during which the domain was active,
the total time between the flushes and the number of queries. For example, in
Figure~\ref{fig:merge_windows}, between flush 0 and
1, Domain A has a \ttt{count} (total number of queries) of $12$ and an \ttt{active\_time} (total time covered by time windows) of $3$, and
Domain B has a \ttt{count} of $5$, and an \ttt{active\_time} of $1$.
If $flush(i)$ is the timestamp of the $i$-th
flush, we define the density at time $i$ as:
\[density(i) = \frac{count}{active\_time}\times \frac{1}{flush(i+1) - flush(i)}.\]
\vspace{-0.1cm}

The first fraction represents the density of requests in the aggregated time
window. The second fraction normalizes this value by the size of the flush window so that all domains have a comparable
 density, as the flushes are not evenly spread.
Therefore, $density(0)$ for domain A is $12/3 \times 1/3 = 4/3$ and $5/1 \times 1/3 = 5/3$ for domain B.

For the \emph{Max Global Variation}, the \emph{flushes} are computed using the time windows of all domains in our ground truth (the numbered flushes in Figure~\ref{fig:merge_windows}). This results in fewer \emph{flushes} but the traffic density between different domains can be compared (as they all use the same time steps).
The \emph{Max Local Variation} of a domain is computed using only the time windows of this domain to compute the \emph{flushes} (numbered \emph{flushes} plus domain \emph{flushes} in Figure~\ref{fig:merge_windows}). The \emph{Local Max Variation} uses more time steps so the density is more precise, but these time steps are different for each domain and have a tendency to reduce the detection of sudden bursts following a long inactivity window.

\vspace{-0.2cm}
\section{Classifier Metrics and Algorithms}\label{app:classifier_metrics}
The performance of each classifier is measured with three metrics: 
\begin{description}
    \item[F1-score:] $\frac{2TP}{2TP + FP + FN}$, with $TP, FP$ and $FN$ being respectively the number of True Positives, False Positives, False Negatives
    \item[True Positive Rate (TPR):] $\frac{TP}{TP + FN}$: proportion of spam domains accurately flagged as spam.
    \item[True Negative Rate (TNR):] $\frac{TN}{TN + FP}$: proportion of benign domains accurately flagged as benign. 
\end{description}

To calculate performance metrics, we use the $k$-fold technique: the whole ground truth dataset is
split in 5 equal parts. 
We select one fold for testing and
train the model using the $k-1$ remaining folds.
We repeat this process for each fold. 
Each metric is the average of the five iterations.

\section{Dataset Statistics\label{sec:datastats}}

Table~\ref{fig:dataset_stats} shows the number of queries and unique domains at each data collection and analysis stage. The first step captures DNS \ttt{TXT} queries to newly registered domain names observed in the passive DNS feed. The next step retains only the \ttt{TXT} queries that contain valid SPF data. Then, we build ground truth with the approach described in Section~\ref{sec:ground_truth}.
\begin{table}[h!]
\caption{Number of queries and unique domains in the dataset at different stages}
\label{fig:dataset_stats}
\begin{center}
\begin{tabular}{l@{\hskip 0.2in}c@{\hskip 0.2in}c@{\hskip 0.2in}c}
Stage & Queries & Unique domains & Spam domains \\
\midrule
1. Traffic to new domains & 399M &  14M & 0.8\% \\
2. SPF traffic & 36M & 1.4M & 1.5\% \\
3. Ground truth & 26M & 40,224 & 5.9\% \\
\end{tabular}
\end{center}
\end{table}

\section{Classification Results}

\begin{table}
\caption{Classification results for the Random Forest classifier on the ground truth dataset.}
\label{fig:classification_results}
\begin{center}
\begin{tabular}{c|c|c|c}
    \diagbox{Our method}{Blacklists} & Spam & Benign & \textbf{Total}\\
    \hline
    Spam & TP = 1 716 & FP = 210 & \textbf{1 926} \\
    \hline
    Benign & FN = 676 & TN = 37 622& \textbf{38 298} \\
    \hline
    \textbf{Total} & \textbf{2 392} & \textbf{37 832} & \textbf{40 224}\\
    \hline \hline 
     & \textbf{TPR} & \textbf{TNR} & \textbf{F1-score} \\
     & 71.7\% & 99.4\% & 79.5\% \\
\end{tabular}
\end{center}
\end{table}

Table~\ref{fig:classification_results} shows the results of the Random Forest classifier using static and dynamic features (SPF Rules, SPF Graph and Time Analysis features). 
It corresponds to the model from Figure~\ref{fig:roc} with a TPR of 0.717 and FPR of 0.006. 
The second and third columns (Spam and Benign) represent how commercial blacklists (SpamHaus and SURBL)
classified the domains (ground truth data), whereas the second and third row represent how our system classified the same domains. 
For example, in the table we can note that 676 domains were classified as Benign by our classifier, but they appear in the commercial blacklists---this represents the number of False Negatives (FN). 
The second part of the table shows the metrics used to evaluate our classifier (TPR, TNR, and F1-score) as described in Appendix~\ref{app:classifier_metrics}.

\end{document}